\documentclass[9pt, conference]{IEEEtran}
\IEEEoverridecommandlockouts

\usepackage{cite}
\usepackage{amsmath,amssymb,amsfonts}
\usepackage{algorithmic}
\usepackage{graphicx}
\usepackage{textcomp}
\usepackage{xcolor}
\usepackage{array} 
\usepackage{makecell} 
\usepackage{booktabs}
\usepackage{enumitem}
\usepackage[T1]{fontenc} 

\newcommand\df{\textbf{DFN}}
\newcommand\dfing{\textbf{DFiN}}
\newcommand\dfopt{\textbf{DFiN-Sel}}
\newcommand\dfse{\textbf{DFiN-SE}}
\newcommand\dfsi{\textbf{DFiN-SI}}
\newcommand\dfcnn{\textbf{DFiN-Cnn14}}
\newcommand\dfatt{\textbf{DFiN-XAtt}}

\definecolor{darkred}{rgb}{0.6, 0, 0}

\usepackage[acronym, shortcuts, nohypertypes={acronym}]{glossaries}
\newacronym{DNN}{DNN}{deep neural networks}
\newacronym{CNN}{CNN}{convolutional neural network}
\newacronym{SE}{SE}{speech enhancement}
\newacronym{DFN}{DFN}{DeepFilterNet}
\newacronym{DFiN}{DFiN}{DeepFingerNet}
\newacronym{HA}{HA}{hearing aid}
\newacronym{ERB}{ERB}{equivalent rectangular bandwidth}
\newacronym{STFT}{STFT}{Short-Time Fourier Transform}
\newacronym{DF}{DF}{Deep Filtering}
\newacronym{MF}{MF}{multi-frame filterbank}
\newacronym{STOI}{STOI}{Short-Time Objective Intelligibility}
\newacronym{PESQ}{PESQ}{Perceptual Evaluation of Speech Quality}
\newacronym{SDR}{SI-SDR}{Scale-Invariant Signal-Distortion Ration}
\newacronym{MOS}{MOS}{Mean Opinion scores}

\usepackage[capitalize]{cleveref}
\usepackage{footmisc}

\def\BibTeX{{\rm B\kern-.05em{\sc i\kern-.025em b}\kern-.08em
    T\kern-.1667em\lower.7ex\hbox{E}\kern-.125emX}}
\title{DFingerNet: Noise-Adaptive Speech Enhancement for \\ Hearing Aids}
\author{
    \textit{Iosif Tsangko}$^{*1,2,3}$, \textit{Andreas Triantafyllopoulos}$^{*2,3}$\thanks{$^*$Equal contribution}, \textit{Michael Müller}$^4$, \\ \textit{Hendrik Schröter}$^4$, \textit{Björn W. Schuller}$^{1,2,3,5,6}$ \\
    \\
    $^1$EIHW -- Chair of Embedded Intelligence for Health Care and Wellbeing, University of Augsburg, Germany \\
    $^2$CHI -- Chair of Health Informatics, Technical University of Munich, Germany \\
    $^3$MCML -- Munich Center for Machine Learning, Munich, Germany \\
    $^4$WS Audiology, Research and Development, Erlangen, Germany \\
    $^5$GLAM -- Group on Language, Audio, \& Music, Imperial College London, UK \\
    $^6$MDSI -- Munich Data Science Institute, Munich, Germany \\
    
    \texttt{iosif.tsangko@tum.de}}

\begin{document}

\maketitle

\begin{abstract}
The \textbf{DeepFilterNet} (\df{}) architecture was recently proposed as a deep learning model suited for hearing aid devices.
Despite its competitive performance on numerous benchmarks, it still follows a `one-size-fits-all' approach, which aims to train a single, monolithic architecture that generalises across different noises and environments.
However, its limited size and computation budget can hamper its generalisability.
Recent work has shown that in-context adaptation can improve performance by conditioning the denoising process on additional information extracted from background recordings to mitigate this.
These recordings can be offloaded outside the hearing aid, thus improving performance while adding minimal computational overhead.
We introduce these principles to the \df{} model, thus proposing the \textbf{DFingerNet} (\dfing{}) model, which shows superior performance on various benchmarks inspired by the DNS Challenge.
\end{abstract}


\section{Introduction}
\label{sec:intro}
\Acp{HA} 
enhance auditory perception and quality of life among those with hearing impairments~\cite{kochkin10-MTE}, particularly in challenging, noisy environments.
They aim to distinguish between speech and noise, enhancing the former while suppressing the latter, thereby allowing wearers to participate in conversations even in busy settings like restaurants -- the well-known \emph{cocktail party problem}~\cite{cherry1953-SEO}.
Speech enhancement (SE) is arguably one of the core modules of~\ac{HA} devices, with real-time approaches traditionally depending on statistical models~\cite{ephraim1984-SEU}. 
In recent years, data-driven deep learning models trained on large datasets have shown significant promise in \ac{SE} tasks~\cite{pascual17-SPE, milling24-AEF, hu20-DDC, shetu24-ULC, schroter22-LLS}.
While modern \ac{DNN} architectures yield improved results, their computational complexity can challenge the real-time processing capacities of \acp{HA}, with the number of parameters and operations being 
prohibitive for the capabilities of most contemporary \acp{HA}~\cite{luo19-CTS}.
This has led to the introduction of more lightweight \ac{DNN} models~\cite{valin20-PMA, valin18-HDL}, including the \df{} family of models~\cite{schroter22-DF1, schroter23-DPM}, which also offers a version optimised for \ac{HA} devices~\cite{schroter23-DMF}.

However, standard \ac{DNN} SE models struggle to adapt dynamically to varied acoustic environments because they are statically trained and deployed without considering changing background noise conditions. Such monolithic approaches can hamper generalisation, especially in cases where the mismatch between the training and deployment environments is large~\cite{li21-ASD, wang18-SSS}.
Adaptive methods, such as ones relying on noise ``fingerprints'' have shown promise in improving SE performance and adaptability to new, unseen environments~\cite{keren18-SSE, liu21-NHA}.
These methods rely on capturing knowledge about the background noise during deployment to adapt the denoising process of an SE model~\cite{williamson15-CRM} (similar to the estimation of noise characteristics needed by traditional algorithms like the Wiener filter~\cite{aubreville18-DDH, schroter20-LON}).

Though promising, these methods have only been applied to large SE models whose size is prohibitive for \acp{HA}.
In the present contribution, we aim to bridge this gap by adopting this approach for a contemporary \ac{DNN} architecture optimised for \ac{HA} use, namely \df{}~\cite{schroter23-DMF}.
Specifically, we investigate different architectural choices for injecting information about the background noise (in the form of noise fingerprints).
We further go beyond prior work by investigating whether this noise conditioning can be \emph{selectively} turned on and off without changing the underlying model, and leverage the pretrained weights of an existing model to boost performance.
This results in an optional add-on component meant to enhance the performance of an existing SE model only when desirable -- rather than resulting in a new model which substitutes the old one.
This added flexibility is desirable in the complex \ac{HA} scenario where calibration to the end-user is paramount for user satisfaction.
Finally, we include an analysis of edge-cases for fingerprint conditioning that sheds more light on the workings of this mechanism that have not yet been investigated in prior work.

\begin{figure*}[t]
  \centering
  \includegraphics[width=\textwidth]{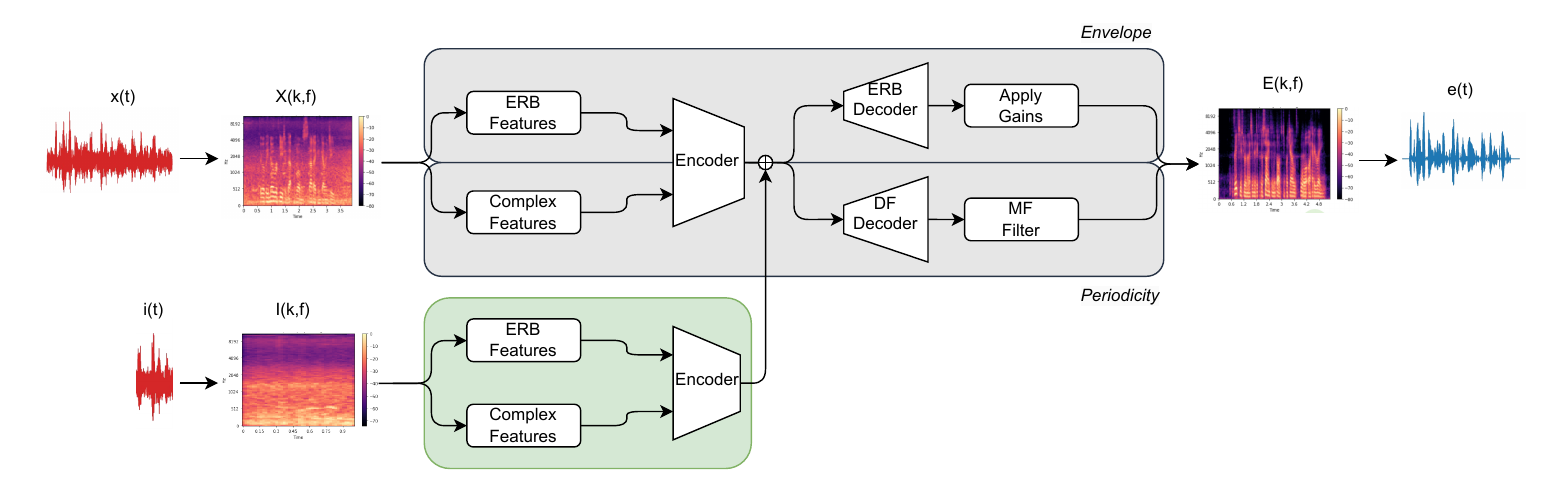}
  \caption{Architecture of the \ac{DFiN}: 
  The model processes input noise and noisy fingerprint through separate encoders for \ac{ERB} and complex features. After fusion (\cref{eq:additive}), the features are decoded by respective \ac{ERB} and \ac{DF} decoders. The decoded \ac{ERB} features are used to apply gains, while the noisy spectrum is filtered using \ac{MF} filters. Combining gains with the filtered spectrum  results in the enhanced spectrum estimate $E$, which is inverted to produce $e(t)$.  
  }
  \label{fig:dfingernet}
\end{figure*}
\section{Methodology}
\label{sec:methodology}
In this section, we describe the methodology used in our study, focusing on the model architecture, the integration of noise-conditioning fingerprints, and our evaluation protocol.

\textbf{Base model:}
The model architecture is based on \df{}, incorporating a proprietary \ac{HA} filter bank for feature extraction.
The \ac{HA}-optimised version of \df{}~\cite{schroter23-DMF} employs a two-stage SE framework. The first stage operates in the \acrfull{ERB} domain to recover the speech envelope, while the second stage uses multi-frame (MF) filtering~\cite{mack19-DFS, huang11-AMF} to enhance the periodic components of speech up to 4 kHz. It employs a 24 kHz uniform polyphase HA filter bank~\cite{bauml08-UPF} instead of the traditional \ac{STFT}, allowing integration into hearing aids. 

\textbf{DFingerNet (DFiN):} 
The conceptual framing of our contribution is as follows.
We assume the presence of additive noise uncorrelated to speech:
\begin{equation}
x(t) = s(t) + n(t) \Rightarrow	X(k, f) = S(k, f) + N(k, f),
\end{equation}
where $X(k,f), S(k, f)$ and $N(k,f)$ are the time-frequency representations of the time domain signals $x(t), s(t)$, $n(t)$ respectively, and $k$, $f$ are the time and frequency indices.
The goal of the original \df{} model is to directly provide an estimate $\tilde{S}$ of the clean spectrum $S$ given the noisy spectrum $X$.

\textbf{Our main goal is to enhance denoising by utilising a \emph{fingerprint} recording of the noise environment, $i(t)$.}
Importantly, $i(t)$ is \textbf{not identical} $n(t)$, but is recorded in the same environment as $n(t)$, thus sharing the same \emph{noise profile}.
By injecting this additional information into the main denoising model, we aim to improve its denoising capabilities.

In detail, the encoder used by Schröter et al.~\cite{schroter22-DF1, schroter22-DF2} processes the input $X(k, f)$ in two separate streams, producing ERB and complex features $X_{erb}(k,f), X_{df}(k,f)$, which are combined into an embedding (\cref{fig:dfingernet}, gray box). We call this encoder 
$\mathcal{F}_{main}(\cdot)$ and its output $\mathcal{E}_{main}(k, f')$. 
Our model includes a new module: the fingerprint encoder $\mathcal{F}_{fing}(\cdot)$ (\cref{fig:dfingernet}, green box).
It processes the fingerprint spectra 
$\mathcal{I}(k, f)$, where $k$ denotes the time axis of the fingerprint and $f$ its frequency axis (which are, in general, different to the time axis of the main input). 
Consequently, we have the following encoded embeddings:
\begin{equation}
\mathcal{E}_{main}(k, f') = \mathcal{F}_{\text{enc}}\left(X_{\text{erb}}(k, f), X_{\text{df}}(k, f)\right)
\label{eq:main_encoder}
\end{equation}
\begin{equation}
\mathcal{E}_{fing}(k, f') = \mathcal{F}_{fing}\left(\mathcal{I}_{\text{erb}}(k, f), \mathcal{I}_{\text{df}}(k, f)\right),
\label{eq:fing_encoder}
\end{equation}
where, for simplicity, we assume that the feature dimension of $\mathcal{F}_{fing}(\cdot)$ is identical to the feature dimension of $\mathcal{F}_{main}(\cdot)$ (in practice, this alignment can be achieved straightforwardly using a linear projection layer).
Finally, we fuse the two embeddings and derive the final one which will be processed further. 
The process is as follows:
\begin{equation}
\mathcal{E}(k, f') = g(\mathcal{E}_{main}(k, f'), \mathcal{E}_{fing}(k', f'))\text{,}
\label{eq:fusion}
\end{equation}
where $g$ is a fusion function.
In the simplest case, $g$ takes the form of additive fusion of the $\mathcal{E}_{fing}$ embeddings, averaged 
over time, or concretely:
\begin{equation}
    \mathcal{E}(k, f') = \mathcal{E}_{main}(k, f') + \frac{1}{K}\sum_{l \leq K}\mathcal{E}_{fing}(l, f')\text{,}
    \label{eq:additive}
\end{equation}
where $K$ is the total duration of the fingerprint (in frames).
As discussed below, we additionally experiment with attention-based fusion of the two time-series using multihead attention.
Finally, we process this embedding with the \df{} decoder~\cite{schroter22-DF1} to estimate the enhanced audio signal.
\textbf{Model variants:} We explore several model variants to assess their performance under different conditions:
\begin{description}[labelwidth=0em, labelsep=0.2em, leftmargin=1em]
    \item[\dfing{}:] the simplest version of our model, where the fingerprint encoder architecture is identical to that of the main network, but randomly initialised. 
    \item[\dfsi{}:] same as above, but the fingerprint encoder is initialised to the same pretrained state as the main encoder, but then trained independently. 
    \item[\dfse{}:] fingerprint encoder weights are additionally coupled to those of the main encoder, i.\,e., the same encoder is applied to both inputs. This has the advantage of further decreasing the memory footprint and optionally allowing the fingerprint encoder to be run on the \ac{HA} device itself.
    \item[\dfcnn{}:] uses a pretrained Cnn14 model~\cite{kong20-PLS} (trained for audio tagging on AudioSet) instead of a \textbf{DFN}-like encoder. 
    \item[\dfatt{}:] the additive fusion is replaced by cross-attention (where we use $\mathcal{E}_{main}$ as query and $\mathcal{E}_{fing}$ as keys and values~\cite{vaswani17-AIA}). This aligns the frames in $\mathcal{E}_{fing}$ with $\mathcal{E}_{main}$ (conceptually similar to time warping) by summarising the embedding before adding them to $\mathcal{E}_{main}$. We use a multihead attention with 4 heads, followed by a feed-forward layer and adding a residual connection.
\end{description}

We note that the main encoder $\mathcal{F}_{main}(\cdot)$ is always initialised from the original pretrained state of \cite{schroter23-DMF}. In preliminary experiments, this was found to be always beneficial compared to training from scratch and allows us to leverage the longer training of the \df{} model; the same is true for the decoder.
Moreover, all the model variants except \dfatt{} feature the simple additive fusion described in \cref{eq:additive}.

\textbf{Training dataset:} Addressing the need for robust evaluation of speech enhancement models, this study utilises diverse datasets and innovative mixing techniques to simulate real-world scenarios.
We adapted the mixing scripts of the DNS dataset~\cite{reddy20-TID} to mix only one speech signal with only one noise sample while additionally cropping the initial 1\,s of the noise sample to be used as the fingerprint, similar to \cite{keren18-SSE} (code for the dataset creation will be released after the review period).
Thus, each sample in our training dataset consists of a clean speech signal, a noise signal split into the fingerprint (first 1\,s) and noise to be used for mixing (rest), and the mixture (generated with a random SNR in the range [-5, 20]).
In our approach, we utilised the entire speech dataset for the training process to ensure the model was exposed to a diverse set of vocal inputs. However, for noise sampling, we selectively drew from the AudioSet~\cite{gemmeke17-ASA} dataset. This strategy was intentionally adopted to reserve other noise datasets, specifically ESC-50, FSD50K, and DEMAND, for evaluation purposes. By doing so, we maintained a clear separation between training and evaluation noise sources, enabling a more rigorous and unbiased assessment of model performance during testing.

\textbf{Hyperparameters:} We use the same set of hyperparameters, including learning rate, batch size, and number of layers, as the DFN model~\cite{schroter23-DMF}.
All models are fine-tuned for 30 epochs using a dataset of 50,000 samples.

\textbf{Evaluation datasets\footnote{https://github.com/ATriantafyllopoulos/dns-custom-mixtures}:} For a comprehensive evaluation, we used the VoiceBank corpus (VCTK)~\cite{veaux13-TVB} for clean speech and combined it with noise from the DEMAND~\cite{thiemann13-TDE}, FSD50k~\cite{fonseca21-FAA}, and ESC-50~\cite{piczak15-EDF} datasets; we refer to these datasets as VCTK-DEMAND and VCTK-FSD, respectively, in the following sections.
We created new test sets by mixing noise and speech at various signal-to-noise ratios (SNRs), ranging from -15\,dB to 25\,dB, to test the model's robustness across different noise intensities and categories.

Specifically, for the VCTK-FSD test set, we selected the top 20 longest files from the over 200 noise categories in FSD50k and mixed these with random VCTK speech files at SNRs ranging from -5 to 5\,dB.
This approach assesses the system's granularity in handling various noise categories.
We also revisited the classic setup with the VCTK-DEMAND dataset. Here, we mixed original VoiceBank speech with noise captured from the last seconds of the recordings used in the original VCTK-DEMAND dataset and examined the robustness to fingerprint mismatches by using noise fingerprints from \emph{different} past time points, ranging from 3 to 120 seconds earlier~(\cref{fig:demand}). These datasets and mixing strategies test the model's adaptability to different noise types and intensities, rigorously evaluating its real-world effectiveness.

We also utilised all `fold-0' files from ESC-50, encompassing 400 files across 50 noise categories, to create a test set called VCTK-ESC. Each noise file was mixed with a random VCTK speech sample, targeting comprehensive evaluation against different SNRs.

\begin{table}[t]
  \centering
    \caption{
        Performance comparison of the \df{}, \dfing{}, \dfse{}, \dfsi{}, \dfcnn{}, and \dfatt{} models on the VCTK-FSD dataset.
        The top row shows metrics computed on noisy speech before denoising; the bottom part of the table shows improvement of enhanced over noisy speech.
    }
  \label{tab:initialisation_strategies}
  \begin{tabular}{lcccc}
    \toprule
    \textbf{Model} & \textbf{SI-SDR} ($\uparrow$) & \textbf{PESQ} ($\uparrow$) & \textbf{STOI} ($\uparrow$) & \textbf{PMOS} ($\uparrow$)\\
    \midrule
    \emph{Mixture} & -4.70 & 1.21 & 0.70 & 2.34\\ 
    \midrule
    \midrule
    \multicolumn{5}{c}{\emph{Improvement over noisy mixture ($\Delta$)}}\\
    \midrule
    \df{} & 10.84 & 0.31 & 0.05 & 0.48\\
    \dfing{} & \textbf{11.35} & 0.39 & 0.07 & 0.54\\
    \dfse{} & 11.05 & 0.39 & 0.06 & 0.53\\
    \dfsi{} & 9.36 & 0.25 & 0.05 & 0.49\\
    \dfcnn{} & 11.23 & 0.40 & 0.07 & 0.55\\
    \dfatt{} & 11.31 & 0.36 & 0.07 & 0.55\\
    \bottomrule
  \end{tabular}
\end{table}


\textbf{Evaluation metrics:} To evaluate the performance of the models, we use several standard metrics: \ac{SDR}~\cite{le19-SHB}, \ac{STOI}~\cite{taal11-AAF} which measures speech intelligibility, and \ac{PESQ}~\cite{rix01-PES}; for each of them, we compute the difference ($\Delta$) between noisy and enhanced speech.
We also consider DNSMOS~\cite{reddy21-DAN} to estimate machine-learning-based \ac{MOS}, specifically focusing on the P808 Model, which predicts human listening experience based on ITU-T P.808 ratings.



\textbf{Selective Fingerprint Use:} As mentioned, we are interested in making the use of fingerprints \emph{optional} in order to accumulate scenarios where they cannot be reliably obtained.
To evaluate the model's robustness in such cases, we disable fingerprints during inference by setting $\mathcal{E} = \mathcal{E}_{\text{main}}$ in \cref{eq:fusion}.

\begin{table}[t]
    \caption{
    Comparison of \dfing{} with \dfopt{} on the VCTK-FSD dataset illustrating the superiority of training with randomly deactivate embeddings when tested both with and without noise fingerprints.
    }
    \label{tab:selective}
    \centering
    \begin{tabular}{c|ccc}
    \toprule
        \textbf{Model} & \textbf{$\Delta$\text{SI-SDR}} (dB) & \textbf{$\Delta$\text{PESQ}} & \textbf{$\Delta$\text{STOI}} \\
        \midrule
        \multicolumn{4}{c}{\emph{Without Fingerprints}} \\
        \midrule
        \dfing{} & 10.63 & 0.36 & \textbf{0.07} \\
        \dfopt{} & \textbf{11.11} & \textbf{0.40} & \textbf{0.07} \\
        \midrule
        \multicolumn{4}{c}{\emph{With Fingerprints}}\\
        \midrule
        \dfing{} & \textbf{11.35} & 0.39 & \textbf{0.07} \\
        \dfopt{} & 11.34 & \textbf{0.42} & \textbf{0.07} \\
        \bottomrule
    \end{tabular}
\end{table}
To prevent a mismatch between training and inference, we also trained one additional model where fingerprints were used in a fraction $p$ of the training iterations and were omitted in the remaining $1 - p$ iterations.
Specifically, we experimented with $p = 0.5$ ($50\%$ of batches with fingerprints).
To counterbalance the fact that the fingerprint encoder sees $50\%$ less gradient updates in this setup, we additionally doubled the number of training iterations to $60$ epochs.
We denote this model as \dfopt{}.

\textbf{Stress tests:} Finally, in order to estimate upper and lower bounds of performance when using fingerprints, we perform the following stress tests: a) we set fingerprints identical to the speech that needs to be enhanced (lower bound); this adversarial input establishes the worst-case scenario where the noise that needs to be removed is equated with the speech that needs to be preserved; and b) we set fingerprints identical to the noise that needs to be removed, thus giving the complete information to the model (upper bound).

\begin{table}[t]
  \centering
  \caption{
  Stress tests for our proposed \dfing{} model on the VCTK-FSD dataset when using the ground truth clean (lower bound) or noisy (upper bound) signals as fingerprints.
  }
  \label{tab:stress}
  \begin{tabular}{lccc}
    \toprule
    \textbf{Fingerprints} & \textbf{$\Delta$\text{SI-SDR}} (dB) & \textbf{$\Delta$\text{PESQ}} & \textbf{$\Delta$\text{STOI}} \\
    \midrule
    Clean speech & 10.61 & 0.36 & 0.06 \\
    Noise signal & 11.46 & 0.39 & 0.07 \\
    \bottomrule
  \end{tabular}
\end{table}
\section{Results}
\label{sec:res}
Our main results are shown in \cref{tab:initialisation_strategies}, which provides a performance comparison of the \df{}, \dfing{}, \dfse{}, \dfsi{}, \dfcnn{}, and \dfatt{} models on the VCTK-FSD dataset.
The top row shows metrics computed before denoising; on average, our data features an SI-SDR of $-4.70$\,dB, showcasing the challenging nature of the dataset, with PESQ being $1.21$ and STOI $0.70$, indicating that subjective perceptual quality and intelligibility are also hampered.
Enhancement results indicate that \dfing{} achieves the highest scores in most metrics, with a $\Delta$SI-SDR of $11.35$\,dB, $\Delta$PESQ of $0.39$, and $\Delta$STOI of $0.07$.
Its superiority over the baseline \df{} is further shown when computing non-intrusive ITU-T P.808 scores, with \dfing{} scoring $2.88$ and \df{} $2.83$.

\begin{figure}[t]
\centering 
\includegraphics[width=.9\linewidth]{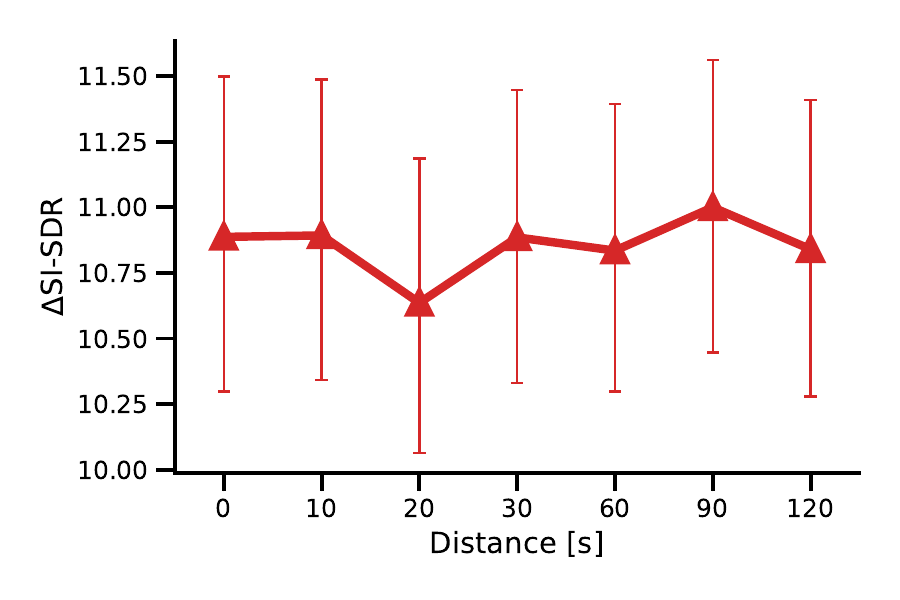}
\caption{Stability analysis of the \dfing{} model on the VCTK-DEMAND dataset, showing \textbf{$\Delta$SI-SDR} as a function of the time difference (in seconds) between the noise mixture and noise fingerprints.}
  \label{fig:demand}
\end{figure}




Random initialisation of the fingerprint encoder yields better performance than initialising with the same weights as the main encoder (\dfsi{}); this is expected as the fingerprint encoder is meant to focus on properties of the sound, whereas the main encoder has been trained to distinguish between background noise and speech.
Surprisingly, coupling the two encoders by sharing their weights (\dfse{}) recovers some of the lost performance; this counterintuitive observation warrants further scrutiny in follow-up work.
We additionally note that using a model pretrained for audio tagging on AudioSet (\dfcnn{}) results in marginally worse performance; while this could be a side-effect of the particular architecture we used, it suggests that audio tagging alone is insufficient for improvement, though it still outperforms the baseline.
Finally, the more sophisticated fusion mechanism employed in \dfatt{} does not result in further gains, scoring marginally below \dfing{} despite the added computational overhead (it needs to run on-device) and the fact that it cannot be easily translated into a streaming operator (the softmax operation depends on the elements that are included; we include all frames in our implementation, which makes it non-causal).
Thus, our conclusion is that the simplest possible setup (using an identical, randomly-initialised encoder and additive fusion) shows to be the most effective.

\textbf{Selective fingerprint use}: \cref{tab:selective} shows the performance of \dfing{} when fingerprints are deactivated during inference (the addition of embeddings is omitted).
Performance degrades to a $\Delta$SI-SDR of $10.63$\,dB, lower than the baseline performance of \df{} ($10.84$\,dB), suggesting that naively disabling fingerprints during inference can be harmful.
However, when this selective disabling is also used during training, as in \dfopt{}, performance degrades more moderately ($11.11$\,dB)
while the model retains an almost identical result with fingerprints actively used ($11.34$\,dB).

\textbf{Stress tests:} \cref{tab:stress} shows the behaviour of \dfing{} under two extreme adaptation scenarios. Using the original clean speech signal leads to a substantial performance drop ($10.61$\,dB), simulating the case where erroneous fingerprints occur 
(e.\,g., due to an underperforming voice activity detection module which identifies background noise). Conversely, supplying the noise signal used for the mixing, results in a slight performance boost ($11.46$\,dB).


\textbf{Robustness to distribution shift:} 
The \dfing{} model was additionally evaluated using the DEMAND dataset across various conditioning scenarios, as shown in ~\cref{fig:demand}. It illustrates the \textbf{$\Delta$\text{SI-SDR}} performance of \dfing{} when using fingerprints from different parts of the background noise environment, taken at increasing distances from the section of the noise file used to generate the mixture. In practice, we observe minimal degradation even when fingerprints are taken up to 2 minutes before the noisy mixture is created, highlighting that – for relatively stable background noise conditions – fingerprints can be obtained at sparse intervals. These findings suggest that fingerprint embeddings can be extracted on a different device (such as a smartphone or smartwatch) and then streamed to the hearing aid, seamlessly incorporating them into the system with minimal overhead. We note that the  base \df{} model outperformed \dfing{} on this dataset, achieving a $\Delta$\text{SI-SDR} of 11.85. This is likely because \df{}'s simpler architecture is better suited to the consistent noise conditions in DEMAND. In contrast, \dfing{} is optimised for dynamic environments, where its adaptive mechanisms excel.
\begin{figure}[t]
  \centering 
  \includegraphics[width=\linewidth]{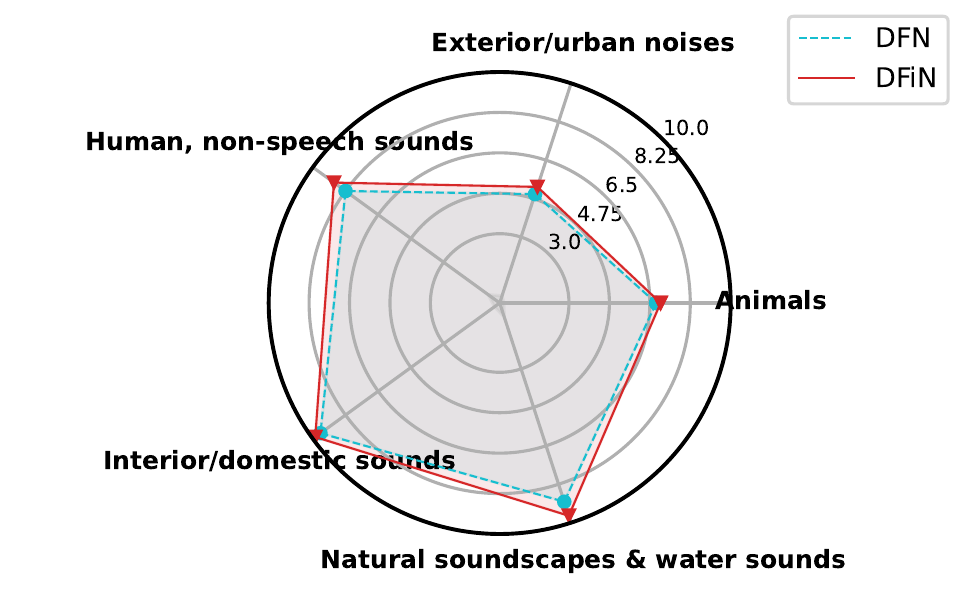}
  \caption{Evaluation on the high-level categories of the classes in the ESC dataset. The radar chart illustrates the \textbf{$\Delta$SI-SDR} performance boost across these high-level categories achieved by our model (\dfing{}) compared to the baseline (\df{}).}
  \label{fig:radar}
\end{figure}

\textbf{Generalisation across noise categories:}
The VCTK-ESC test set allows for a more granular analysis of model performance with respect to different noise categories.
\cref{fig:radar} presents \textbf{$\Delta$SI-SDR} for the 5 high-level categories of ESC-50, with \dfing{} outperforming the baseline \df{} across all high-level categories.
Interestingly, this granular presentation of results allows us to uncover shortcomings of the original \df{} model (which \dfing{} inherits), as performance varies substantially across different categories, with ``exterior/urban noises'' and ``animal sounds'' showing the worst performance.
This may be a side-effect of the training data (e.\,g., an underrepresentation of these categories) and warrants closer investigation in future work.

\section{Conclusion}
In this work, we have introduced the \dfing{} model, which extends the \df{} architecture by adding an adaptation module relying on noise fingerprints to enhance speech clarity.
Our approach demonstrates that noise adaptation can be attached to a pretrained model as an extra, optional step to improve performance.
Importantly, robustness to slow-changing background environments shows that the method adds minimal computational overhead, as the fingerprint encoder can be run off-device, which is vital given the restrictions of hearing aid devices.

Future work could investigate the combination of noise-adaptation with the now standard speaker-adaptation~\cite{sivaraman21-ZSP, kim24-ZST} used to personalise denoising, as well as understand the training dynamics of the original \df{} model to overcome its underperformance on certain kind of noises.

\bibliographystyle{IEEEtran}
\bibliography{strings,refs}

\end{document}